%% file: e2c_excess_paper.tex
\documentclass[10pt,letterpaper]{IEEEtran}

\usepackage{times}
\usepackage{amsmath}
\usepackage[]{graphicx}
\usepackage{algorithm}
\usepackage{algorithmic}
\usepackage{verbatim} 
\usepackage[color]{changebar}
\usepackage{xspace}
\let\labelindent\relax
\usepackage[inline]{enumitem}
\usepackage{url}
\usepackage{array}

\begin{document}

\title{Monitoring in the Clouds: Comparison of ECO$_2$Clouds and EXCESS Monitoring Approaches}

\author{\IEEEauthorblockN{Pavel Skvortsov, Dennis Hoppe, Axel Tenschert, Michael Gienger} \\
\IEEEauthorblockA{HLRS, Universit\"at Stuttgart, Stuttgart, Germany \\
skvortsov@hlrs.de, hoppe@hlrs.de, tenschert@hlrs.de, gienger@hlrs.de
}}

\maketitle

\input{00-Abstract.txt}
\input{01-Introduction.txt}
\input{02-E2C_Approach.txt}
\input{03-EXCESS_Approach.txt}

\input{04-Comparison_and_Analysis.txt}

\input{05-Conclusion.txt}

\newcommand{\BIBdecl}{\setlength{\itemsep}{0.28em}}
\bibliographystyle{IEEEtran}

\bibliography{lib}

\end{document}

%% file: 00-Abstract.txt
\newcommand{\ECO}{ECO$_2$Clouds\xspace}
\newcommand{\EXCESS}{EXCESS\xspace}

\begin{abstract}
With the increasing adoption of private cloud infrastructures by providers and enterprises, the monitoring of
these infrastructures is becoming crucial. The rationale behind monitoring is manifold: reasons include saving energy,
lowering costs, and better maintenance. In the e-Science sector, moreover, the collection of infrastructure and
application-specific data at high resolutions is immanent. In this paper, we present two monitoring approaches
implemented throughout two European projects: \ECO and \EXCESS. The \ECO project aims to minimize $\mathrm{CO_2}$
emissions caused by the execution of applications on the cloud infrastructure. In order to allow for eco-aware
deployment and scheduling of applications,
the \ECO monitoring framework provides the necessary set of metrics on different layers including physical,
virtual and application layer. In turn, the \EXCESS project introduces new energy-aware execution models
that improve energy-efficiency on a software level. Having in-depth knowledge
about the energy consumption and overall behavior of applications on a given infrastructure, subsequent executions
can be optimized to save energy. To achieve this goal, the \EXCESS monitoring framework provides APIs allowing
developers to collect application-specific data in addition to infrastructure data at run-time.
We perform a comparative analysis of both monitoring approaches, and highlighting use cases including
a hybrid approach which benefits from both monitoring solutions.
\end{abstract}


%% file: 01-Introduction.txt
\section{Introduction}
\label{sec:intro}

\noindent
Private cloud infrastructures are being increasingly adopted by providers and enterprises.
They include heterogeneous computing, storage and networking resources, which are utilized
by the variety of applications with diverse requirements. Thus, the monitoring of
these infrastructures is becoming important. The reasons for sophisticated and timely
monitoring include not only maintenance-related technical parameters, but also
the analysis of advanced metrics which indicate how to utilize the infrastructure in order to
save energy, lower costs and reduce emissions. At the same time, cloud infrastructures are often used
in terms of e-Science research~\cite{gray2007}, which deals with
sophisticated computer simulations and Big Data use cases. In such cases, the collection
of infrastructure and application-specific data at high resolution is immanent.

Energy-awareness in cloud computing and HPC and embedded systems has become a priority in
recent years \cite{Valentini13}. Although performance keeps to be the topmost objective for companies,
data centers have shown a high interest in reducing their energy consumption
that continues to increase due to several reasons \cite{basmadjian12,liu09,obaidat12}: reasons include the advancement
of Big Data, Fast Data (i.e., Big Data sets, which have to be processed in real-time), the Internet of Things (IoT),
and the recent trend of employing Graphical
Processing Units (GPUs) in order to speed up execution of compute-intensive scientific applications. 
The need for monitoring in high performance computing, in particular in cloud computing is driven by resource planning
and management, data center management, SLA administration, billing, and performance management~\cite{aceto13}.
Aceto et al.~\cite{aceto13} highlight that energy-efficiency is \emph{``a major driver of monitoring data analysis for planning,
provisioning and management of resources''}.
Since increased energy consumption leads to extra operational costs, reducing energy is a driving factor for
data centers~\cite{Valentini13}. 

In this paper, we present two different monitoring approaches focusing
on energy-efficiency, which were
implemented throughout European projects \ECO and \EXCESS respectively.

The \ECO project aims to minimize $\mathrm{CO_2}$ emissions caused by the execution of applications on the cloud
infrastructure. In order to allow for eco-aware deployment and scheduling of applications,
the \ECO monitoring framework provides the necessary set of metrics on different layers including physical,
virtual and application layers. The basic feature allowing for minimizing the
$\mathrm{CO_2}$ emissions is the detailed mix of energy sources which cloud provider sites
EPCC (Edinburgh Parallel Computing Centre), INRIA (French Institute for Research in
Computer Science and Automation) and HLRS (High Performance Computing Center Stuttgart)
are able to support. In addition to the installed power distribution units (PDUs) measuring
the power consumption of physical nodes, the emissions size was computed.
The difference between the energy mix of each site, as well as the
heterogeneity of the computing resources within sites, allows for such
a scheduling of applications execution that the overall carbon footprint (i.e., $\mathrm{CO_2}$ emissions)
of an application is minimized.

Another implementation of a monitoring framework aiming at energy efficiency
is proposed in the EU-funded project \EXCESS (Execution Models
for Energy-Efficient Computing Systems)~\cite{koller14}.
Optimizing the energy consumption on an application level requires a deep understanding of the runtime behavior of applications.
The \EXCESS project introduces new energy-aware execution models that improve energy-efficiency.
The validation and verification of the execution models is needed based on real monitoring data.
The rationale behind the monitoring solution developed in the \EXCESS project is to
raise energy-awareness relevant for software developers. Having in-depth
knowledge about the energy consumption and overall behavior of their applications on a given infrastructure,
subsequent executions can be optimized to save energy. To achieve this goal, the \EXCESS monitoring framework
provides APIs allowing developers to collect application-specific data in addition to infrastructure data at run-time.
The \EXCESS monitoring framework provides \emph{fine-granular} monitoring information
without introducing extra costs on performance that interfere with the application.
EXCESS is concerned with a holistic analysis of systems including the application,
system software and hardware stack in order to detect preventable energy dissipation.
The \EXCESS monitoring framework enables improving the energy-efficiency in high performance computing
and embedded systems, as well as in cloud computing.


In this paper, we perform a comparative analysis of the \ECO and \EXCESS monitoring approaches
We highlight the target use cases including a hybrid approach which benefits from both monitoring solutions.

The rest of this paper is structured as follows: we make an overview of the related work in Section II.
In Sections III and IV, we describe the \ECO and \EXCESS approaches, respectively.
In Section V, we compare the presented approaches and discuss their suitable use cases. Finally, we conclude the paper in Section VI.

\section{Related Work}
\label{sec:rw}

\noindent
Next, we provide a brief overview about existing frameworks for monitoring in computing systems.

Existing systems for monitoring can be distinguished along three dimensions: agent-less, agent-based, and hybrid~\cite{uptime14}.
Agent-based monitoring systems have a client-server architecture. Prior to monitoring a component, agents,
often light-weight applications, are installed remotely on target computers. Each agent then collects specific metric data
such as the CPU usage over time. Periodically, collected data is sent to a monitoring server for storage and analysis.
This approach is crucial for supporting custom, non-standard metrics such as the amount of time a server needs
to respond to a request. Agent-less monitoring systems, by contrast, do not deploy agents on target computers to collect
data; the data is rather gathered through remote calls, e.g., via the Simple Network Management Protocol (SNMP). As
a consequence, this monitoring strategy is limited to basic hardware information available on all platforms by default.
The advantage of such an approach is that no initial software has to be deployed on target computers and the following
low impact on system performance at run-time. Hybrid solutions, finally, support both agent-based and agent-less
monitoring. The ATOM framework of the \EXCESS project
follows an agent-based architecture in order to support application-specific metrics. In turn, the \ECO monitoring
framework relies on the Zabbix-based agents.

Aside from the architecture, monitoring frameworks can be classified via the following nine key properties: scalability,
non-intrusiveness, timeliness, granularity, extensibility, data storage, visualization, adaptability, and
predictability and advanced metrics~\cite{aceto13,katsaros11,telesca14}. The meanings of these properties are listed below.


\begin{itemize}
\item Architecture:
agent-based (producer-consumer principle), agent-less or hybrid;

\item Non-Intrusiveness:
low to very low impact on the system at run-time;

\item Scalability:
high update frequencies of metrics while being low-intrusive; 

\item Timeliness: near-real time update rates of metrics; allowing for “snapshots” at a given point in time;

\item Granularity: profile different levels of granularity (e.g., functions);

\item Extensibility: provide a plug-in system to add extra metrics;
enable application-specific metric support; language-independent interface to transfer data;

\item Data Storage:
low-intrusive, efficient data storage at run-time; data export using common formats (e.g., CSV or JSON);
allow for filtering the data based on time intervals;

\item Visualization: provide basic visualization functionality;

\item Adaptability: enable/disable profiling of specific components configuration of plug-ins at run-time;

\item Predictability: predictors via extensions (e.g., $\mathrm{CO_2}$ prediction);

\item Metrics: non-standard metrics need to be supported;
external measurements to increase data accuracy; profiling of specialized hardware.
\end{itemize}

These demands, in particular the latter one, are not easy to be met by existing monitoring frameworks
such as Ganglia, Lattice, Nagios, Open-NMS, Zenoss, or Zabbix.
For example, key requirements that need to be satisfied by a monitoring framework in case of the \EXCESS project include
supporting different granularity levels to be profiled, low-intrusiveness of the monitoring, and the possibility of
code instrumentation. In case of the \ECO project, it is required to support non-standard $\mathrm{CO_2}$-related metrics,
predict the future energy usage and emissions and to provide a data storage for historic monitoring information.


%% file: 02-E2C_Approach.txt
\section{ECO$_2$Clouds Approach}
\label{sec:e2c_approach}

\noindent
The major goal of the ECO$_2$Clouds project is to enable such eco-aware automated
deployment and execution of applications in the cloud,
which optimizes the $\mathrm{CO_2}$ emissions caused by the physical infrastructure.
In order to achieve this goal, the ECO$_2$Clouds project introduces an approach to monitor the eco-metrics
of the running applications on the different layers
and performs the applications' scheduling based on the retrieved monitoring values.
Next, we describe the monitoring approach developed in the ECO$_2$Clouds project.
For more details about the scheduling used in ECO$_2$Clouds, we refer the interested reader to the work of Wajid et~al.~\cite{Wajid2015}.

\subsection{Architecture of the \ECO Monitoring Approach}

\begin{figure}
  \begin{center}
    \includegraphics[width=0.5\textwidth]{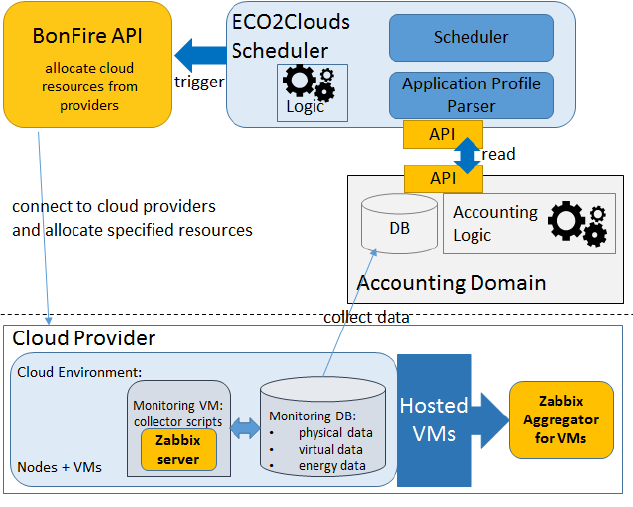}
    \caption{Architecture of the ECO$_2$Clouds monitoring approach \cite{TenschertSG14}}
    \label{fig:E2C_structure}
  \end{center}
\end{figure}

The is \ECO monitoring system is based on the monitoring framework for federated clouds
which was developed during the course of the EU-funded BonFIRE project \cite{HazmiCM12}.
The general architecture of the \ECO monitoring system and its relation
to the other project components is depicted in Figure~\ref{fig:E2C_structure} \cite{TenschertSG14}.
On the cloud provider side, a Zabbix server~\cite{Zabbix} running on a dedicated virtual machine (VM)
collects the monitoring information from the Zabbix agents installed on physical nodes.
The system monitors the metrics from (a) the physical infrastructure, (b) VMs, and (c) applications.
The Zabbix monitoring data is gathered and structured in an Accounting SQL database by
the Accounting Service which runs on a separate VM.
The Accounting DB allows for various combinations of desired experiment,
physical host, provider site, VM, and time frame to be specified in the selection queries. 
The monitoring data from the Accounting DB is then used by the \ECO
Scheduler and Application Controller to provide eco-aware deployment of applications~\cite{Wajid2015}.
On the highest level, user manages the experiments (i.e., applications) and the cloud resources
by using the RESTful BonFIRE API which is based on HTTP commands~\cite{bonfireAPI}.

\subsection{\ECO Technology for Metrics Collection}

The underlying ECO$_2$Clouds solution which enables to consider the eco-aware metrics
is equipment of physical hosts with power distribution units (PDUs).
PDUs provide monitoring systems with exact values of power consumption measurements,
which are later transformed into the carbon footprint (i.e., $\mathrm{CO_2}$ emissions).
The consequent calculation of carbon footprint based on power consumption parameters is possible, since
the distribution of power sources are known for each cloud infrastructure provider.

\begin{figure}
  \begin{center}
    \includegraphics[width=0.5\textwidth]{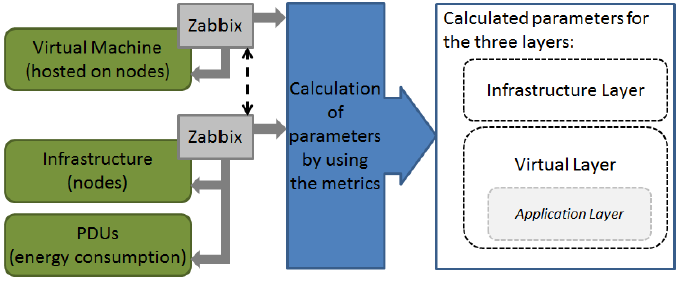}
    \caption{Aggregation and calculation of metrics in ECO$_2$Clouds \cite{TenschertSG14}}
    \label{fig:E2C_combining}
  \end{center}
\end{figure}

Zabbix clients are installed on each physical node and VMs. They collect the pre-defined metrics
including the usual metrics such as memory utilization, I/O throughput and CPU utilization for VMs
as well as PDU-based metrics for physical nodes (see Figure~\ref{fig:E2C_combining}).
The measurement rate differs from 5 to 30 seconds for various metrics, and
the resulting data set was observed to add up to 100 Mbytes per day, depending on the number of running experiments.
The Zabbix server (i.e., metrics aggregator) installed on a dedicated VM (one per each experiment) gathers all
the metrics information from the Zabbix clients within a given experiment.

Different hardware configurations and different conditions regarding the used energy
mix of the providers (EPCC, INRIA and HLRS) were challenging.
In order to adapt the monitoring metrics implementation at all provider sites,
Zabbix templates, bash, Python, and Ruby scripts were implemented. 

\subsection{List of \ECO Metrics}

One of the major contributions of the \ECO approach is calculation of a VM's power consumption.
It is determined through a formula depending on both physical and VM metrics. The metrics required 
for this include the size of a VM defined through the used memory, the data I/O (i.e., the send and receive activities), the disk activity
(i.e., the read and write operations) and the consumed CPU seconds.
The list of major \ECO metrics is presented in Tables~\ref{e2c1} and~\ref{e2c2}. Each metric is marked to
indicate whether it is also supported by the \EXCESS monitoring framework.
We can see that the VM-related metrics are only supported by the \ECO project,
while \EXCESS approach provides energy metrics on the physical level.
For more details about the \ECO metrics we refer to our previous work~\cite{Wajid2015}

\begin{table}
\caption{\ECO metrics (part 1)}
\label{e2c1}
\small
\begin{center}
\begin{tabular}{|m{1.5cm}|m{4cm}|m{0.5cm}|m{1.3cm}|}
\hline
 Metric and layer & Definition & Unit & Support by MF of EXCESS \\ \hline \hline
Task execution time (app. layer) & Time taken to execute a specific task & s & Yes \\ \hline
Application execution time (app. layer) & Time taken to execute whole application & s & Yes\\ \hline
Power consumption (app. layer) & Power currently consumed by application.
& W & Yes \\ \hline
Response time (app. layer) & Average time taken from user request to service response
& s & No\\ \hline
A-PUE (Application PUE) & Application power usage effectiveness; ratio between
total amount of power (P) required by all VMs of application $i$ and the power used to execute $i$-th application task & Ratio & No \\ \hline
Application Energy Productivity (A-EP) & Application Energy Productivity (A-EP);
ratio between number of executions of tasks hosted by all VMs of $i$-the
application and total energy for execution of the VM & $W^{-1}$ & No \\ \hline
Application Green Efficiency (A-GE) & Share of green energy used to run the $i$-th application
& W & No \\ \hline
CPU usage (virt. layer) & Processor utilization (inside a VM) for running VM
& \% & No \\ \hline
Storage usage (virt. layer) & Storage utilization on corresponding storage device; the ratio between
the used disk space and the allocated disk space & \% & Supportable \\ \hline
\end{tabular}
\end{center}
\end{table}

\begin{table}
\caption{\ECO metrics (part 2)}
\label{e2c2}
\small
\begin{center}
\begin{tabular}{|m{1.5cm}|m{4cm}|m{0.5cm}|m{1.3cm}|}
\hline
Metric and layer & Definition & Unit & Support by MF of EXCESS \\ \hline \hline
I/O usage (virt. layer) & Percentage of process execution time in which the disk is busy with read/write activity & \% & No \\ \hline
Memory usage (virt. layer) & Ratio of memory size used by VM to total memory available & \% & No \\ \hline
Power consumption (virt. layer) & Power currently consumed by the given VM & W & No\\ \hline
Disk IOPS & Assessment of I/O operations of a virtual resource
& $\frac{Ops}{s}$ & No\\ \hline
Site utilization (infr. layer) & Current utilization of a single site; ratio of available cores to total cores.
& \% & No \\ \hline
Storage utilization (infr. layer) & Percentage of frontend storage used
& \% & No \\ \hline
Availability (infr. layer, site) & Shows whether the OpenNebula OCCI server answers the requests
 & Bool & No\\ \hline
PUE (infr. layer) & Energy efficiency of a site; measured as ratio between
total facility power and power used by computing equipment
 & Ratio & No \\ \hline
Power consumption (infr. layer) & Power consumed by given host in given time period & W & Supportable \\ \hline
Disk IOPS (infr. layer) & I/O operations of disk within host & IO/s & No \\ \hline
CPU utilization (Infr. layer) & Average utilization of CPUs inside host
& \% & No \\ \hline
Availability (infr. layer, host) & Availability of host & Bool & No \\ \hline
Green Efficiency Coefficient (GEC) (accounting) & Share of energy consumed by site that
is produced by green energy sources & \% & No \\ \hline
Site Infrastructure Efficiency (SIE) (accounting) & Share of power consumed by information technology
equipment power of total facility power & \% & No \\ \hline
Carbon Usage Effectiveness (CUE) (accounting) & Weighted average of CEFs related to
energy sources used in site, where CEF is Carbon Dioxide Emission Factor taken from literature for
each energy source & Ratio & No \\ \hline
\end{tabular}
\end{center}
\end{table}

%% file: 03-EXCESS_Approach.txt
\section{EXCESS Approach}
\label{sec:excess_approach}

\begin{figure*}[t]
	\centering
	\includegraphics[width=0.8\textwidth]{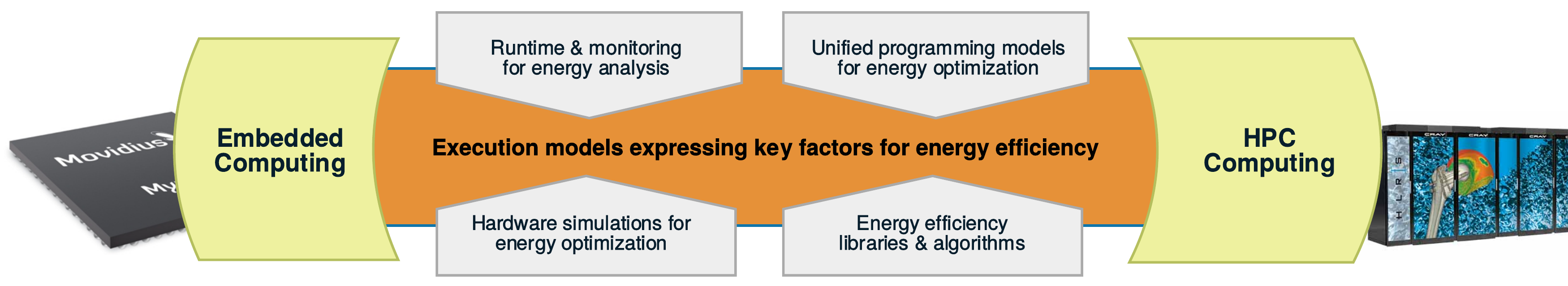}
	\caption{Iterative hardware/software co-design process implemented in the EXCESS project~\cite{koller14}
	}\label{fig:excess-iterative}
\end{figure*}

The key idea behind \EXCESS is a holistic system analysis incorporating high performance and embedded computing~\cite{koller14}.
The iterative hardware/software co-design process implemented in the EXCESS project is illustrated in Figure~\ref{fig:excess-iterative}.
This \EXCESS system's major component is ATOM (neAr-real Time mOnitoring fraMework)~\cite{Hoppe15}. ATOM is a monitoring framework 
which is suitable for both HPC and cloud computing. ATOM tackles the challenge
of analyzing the system's run-time context and overcoming the drawbacks of
existing solutions. ATOM is designed to be
low-intrusive and flexible, while allowing to collect, visualize, and analyze relevant application and infrastructure
data in near-real time. The following items are crucial for optimizing the energy consumption and performance
of applications at run-time, as well as improving energy-awareness in software engineering:

\begin{enumerate*}[label=\itshape\alph*\upshape)]
\item monitoring applications at run-time,
\item collecting large amounts of performance and energy-related data at high frequencies, and
\item providing a user-friendly interface for data analysis.
\end{enumerate*}
The main contributions of ATOM are:
\begin{itemize}
\item Low-intrusive, scalable architecture based on open-source tools and libraries:
\emph{Node.js}, \emph{Elasticsearch}, \emph{D3.js}, and \emph{Kibana}.

\item Flexible, language-independent plug-in system, which supports collecting specific data
including the energy consumption of embedded systems, or connecting to external power measurement devices.

\item Light-weight and easy-to-grasp user library that allows code instrumentation in order to
gather application-specific data not accessible globally.

\item Integration with the PBS resource manager to allow executing and monitoring applications
without any prior knowledge by end users or software developers.

\item Interactive front-end allowing for near-real time exploration of collected data as well as
exporting historical data for further analysis.
\end{itemize}

\begin{figure*}[t]
  \centering
  \includegraphics[width=0.8\textwidth]{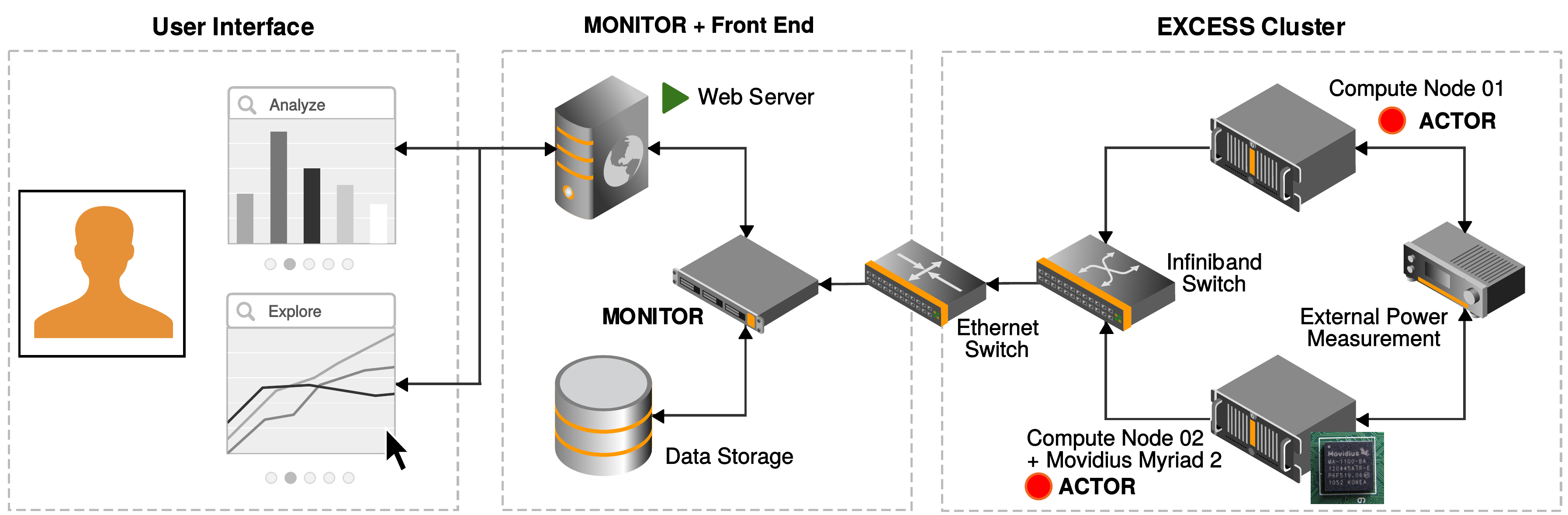}
  \caption{Architecture of the ATOM monitoring framework deployed on the EXCESS cluster
	}\label{fig:monitoring-framework}
\end{figure*}

\subsection{Agent-based Architecture}
ATOM's architecture, as depicted in Figure~\ref{fig:monitoring-framework}, is agent-based. ATOM is composed
of two main components: a server (MONITORing server---\textit{MONITOR}) and multiple, light-weight agents
(Atom CollecTORS---\textit{ACTORS}). The purpose of the ACTORS is to continuously sample node- and
application-specific data and send this information to MONITOR using a user-defined update rate
(e.g., every 100 milliseconds). MONITOR collects the data being sent by each target node and offers
functionalities to query, visualize, and analyze historical data.

The cluster consists of several compute nodes including embedded hardware such as the Movidius Myriad 2 chip.
	When users start a job on the cluster, an ACTOR is started on each node to monitor relevant data while the job
	is running. ACTORS then send sampled data at run-time through a high-bandwidth network (InfiniBand) to the
	MONITOR server. MONITOR manages incoming data, stores it in the database, and also provides a web interface
	accessible by users.

\subsection{ATOM Collectors (ACTORS)}
An ACTOR is deployed on each target node. ACTORS have a set of plug-ins that can be implemented in
any programming language; the only constraint is on the format used for transferring data to MONITOR.
We selected the programming language C for ACTORS and plug-ins. High-level languages such as C++ or
Java impose a performance overhead at run-time, and hence being too intrusive. Since ACTORS are
by default extensible, developers can add plug-ins with ease in order to sample additional metrics.
Plug-ins do not have any limitations on the type of metric to be collected as long as the data is sent
to the server via a pre-defined data format; we selected JSON objects to hold metric data.

\begin{figure}[b]
  \centering
  \includegraphics[width=.48\textwidth]{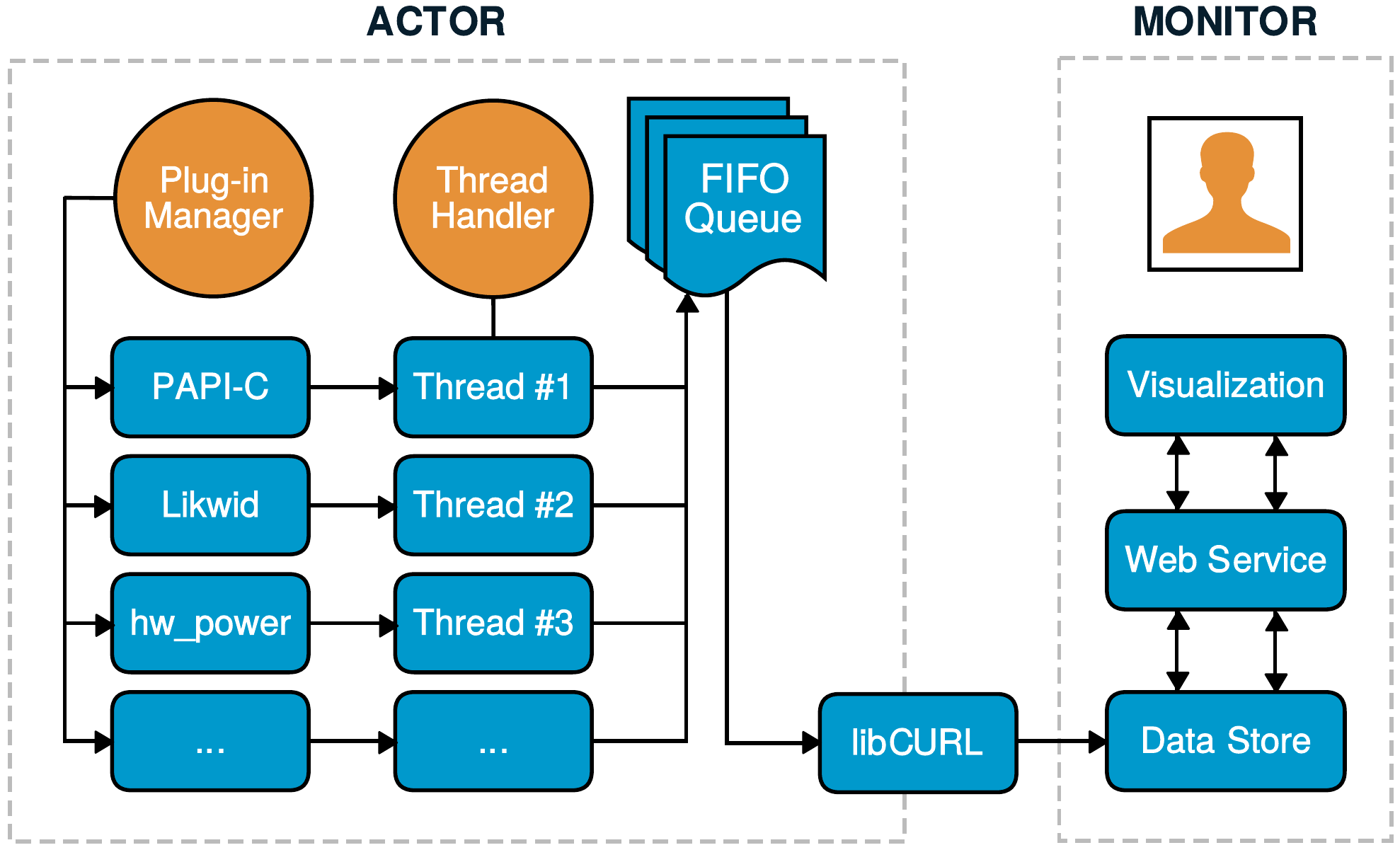}
  \caption{Detailing the interaction between ACTORS (left) and MONITOR (right). Each ACTOR loads at startup
	a list of plug-ins in order to sample relevant metrics. Metric data is buffered in a FIFO queue, before it is
	sent via \emph{libCURL} to the data store. MONITOR is deployed on a separate node in order to reduce the
	performance overhead at run-time.}\label{fig:atom-architecture}
\end{figure}

Figure~\ref{fig:atom-architecture} illustrates the three building blocks of an ACTOR: a plug-in manager,
a thread handler, and a data management component in terms of a FIFO queue connected to \emph{libCURL}.
When ACTORS are started, a default or user-defined configuration file is parsed in order to activate/deactivate
plug-ins, define the relevant metrics to be collected, and to set the sample rate for each plug-in.
During initialization, an \emph{execution ID} is requested from MONITOR; the execution ID is used to
uniquely associate metric data with specific application executions. The thread handling routines then
trigger the plug-in discovery and start a thread for each of the plug-ins using the sample rates previously
defined in the configuration file. In addition to the threads shown in Figure~\ref{fig:atom-architecture},
two extra threads are created: an extra thread aggregates and sends metric data to MONITOR, and another
thread checks the configuration file for changes, which allows sample rates to be changed at run-time.

\subsection{ATOM Monitoring Server (MONITOR)}
The MONITOR server is deployed on a separate node in the EXCESS cluster. As illustrated in
Figure~\ref{fig:atom-architecture}, MONITOR is composed of three building blocks: a data store, a web
service, and a visualization component. As a data storage component, Elasticsearch is used~\cite{elasticsearch14}.
Elasticsearch is a flexible and powerful open source, real-time search and analytics engine. It provides a
distributed, multi-tenant-capable full-text search engine with a RESTful web interface and schema-free JSON
documents. As web server, the server-side JavaScript library \emph{Node.js} is selected~\cite{opennms15}.
Since Node.js is based on Google's V8 engine written primarily in C, Node.js qualifies to be used
for data-intensive real-time applications that run across distributed devices. Basic visualization
functionality is provided via the JavaScript library \emph{D3.js}. This is a popular data visualization
tool~\cite{bostock14}, which is also used by Datameer, a company concerned with visualization of big data~\cite{viau14}. 

The benefits of using these frameworks are manifold: Firstly, the common shared data format JSON eases the
integration of components. Secondly, employing Elasticsearch as (historical) data storage bypasses a
drawback that existing monitoring frameworks such as Nagios or Zabbix impede: They rely on traditional
databases such as MySQL that restrict third-party developers in using pre-defined data models, and cannot
handle vast amounts of data writes and reads in near-real time. With ATOM, we offer developers the flexibility
of selecting their custom JSON objects to store metric data as long as they provide a few mandatory fields
(cf.~Section~\ref{subsec:communication-layer}). In addition, NoSQL databases such as Elasticsearch offer a
feature called auto-sharding, which distributes data automatically and on-demand across multiple nodes in
order to balance query and data load. This feature, among others, makes Elasticsearch an excellent
choice to be used in HPC environments.

\subsection{Communication Layer}\label{subsec:communication-layer}
JSON is selected as the primary data format for data exchange between all ATOM components; the actual
communication between the ACTORS and MONITOR is realized via HTTP. MONITOR exposes a RESTful interface
that each ACTOR can access through simple \emph{CURL} commands.

\subsection{Monitoring Hierarchy}
Since \EXCESS follows a holistic approach to energy reduction, we are interested in profiling different
hierarchy levels of the hardware and software stack: infrastructure, applications, and functions.
Profiling functions, in particular, require code instrumentation. Code instrumentation is hardly
supported by existing monitoring frameworks, because it introduces an additional performance overhead.
Up to date, ATOM supports the following metrics by default: PAPI-C~\cite{terpstra10}, RAPL~\cite{hahnel12},
Likwid~\cite{treibig12}, \texttt{/proc/meminfo}, \texttt{iostat}, and \texttt{hw\_power}~\cite{khabi14}.
It should be noted that metric support will be extended in the future; current plug-ins were selected,
because they represent a minimal set of basic metrics to be considered for profiling.

\paragraph{Infrastructure}
ATOM currently includes the following plug-ins in order to monitor the infrastructure. For performance data,
we rely on PAPI-C and \texttt{/proc/meminfo}. In order to monitor the energy consumption at run-time, we
implemented plug-ins for RAPL and Likwid. An external power measurement system is installed in addition
on the EXCESS cluster (cf.~Figure~\ref{fig:monitoring-framework}) to monitor the energy consumption of
specific components such as the CPU, the GPU, or the whole computational node with high accuracy.

\paragraph{Applications}
In order to profile applications, PAPI-C and \texttt{iostat} can be used to measure performance. Although
PAPI-C is a so-called first-person monitoring tool, i.e.\ it monitors a specific process or thread, we
implemented a third-person variant collecting data on a per-core basis. To the best of our knowledge,
there exists currently no approach to sample the energy consumption of specific applications or smaller
building blocks. We tackle this challenge by monitoring the execution of particular functions by using code instrumentation.

\paragraph{Functions}
Performing a thorough analysis of applications requires to monitor the run-time behavior of functions,
and to collect additional application-specific metric data that is normally not supported by existing
performance or energy measurement tools. Such data includes, for example, the number of user requests
to a web service. As a consequence, we have implemented a light-weight library for source code instrumentation,
which acts like an extra plug-in loaded by an ACTOR. The library has the following key features:

\begin{itemize}
\itemsep1pt\parskip0pt\parsep0pt
\item
sending application-specific data,

\item
profiling specific code fragments (e.g., monitoring execution times of functions), and

\item
retrieving historic metric data for code optimization at run-time.
\end{itemize}

In particular, the last two items are of great interest for EXCESS, because we are interested in optimizing
applications in order to yield a better trade-off between energy and performance at run-time. By profiling
critical code fragments, and being able to directly get feedback about performance and energy consumption,
we can directly use the data as input for optimization of future prediction models.

\subsection{Extensibility}
Custom plug-ins are language-independent due to the following three reasons:
\begin{enumerate*}[label=\itshape\alph*\upshape)]
\item the entire communication is based on HTTP,
\item the data exchange format is JSON, and
\item MONITOR provides a RESTful interface.
\end{enumerate*}
ACTORS are currently written in C, and plug-ins are loaded via shared objects at run-time. The plug-in
for \texttt{iostat} is realized as a Unix shell script using the streaming support of \texttt{awk} to
parse the output of \texttt{iostat} in order to continuously send new metric data to MONITOR.

%% file: 04-Comparison_and_Analysis.txt
\section{Comparison and Analysis}
\label{sec:comparison}

\noindent


Concerning the system architecture, both monitoring frameworks rely on the agent-server architecture.
The monitoring agents called
ACTOR written in C and utilized in \EXCESS, while slower yet more advanced Zabbix agents are used in \ECO.
Next, we consider the major differences between the proposed approaches.

In contrast to the \ECO project, the \EXCESS project is
concerned with a holistic analysis of systems including the application, system software, and hardware
stack in order to detect preventable energy dissipation (cf.~Figure~\ref{fig:excess-iterative}). Thus,
its first goal is not eco-ware automated deployment, but rather enabling software developers to write
energy-efficient applications across different infrastructures. \EXCESS sees a need for a sophisticated
co-design process between software and hardware while developing. 

Optimizing the energy consumption involves a thorough analysis of all layers involved, in particular the
hardware and software layer. From a hardware point of view, \EXCESS develops new energy-saving components
such as the Movidius Myriad 2 platform~\cite{movidius14}. On the software side, \EXCESS evaluates sophisticated solutions with respect to
energy-efficient libraries and algorithms. 

The \EXCESS approach provides near-real time monitoring: the maximal possible frequency of updates achieves 10 Hz,
while the Zabbix system of the \ECO monitoring framework updates data only once per 5 seconds for the most metrics.
As our experiments have shown, the CPU performance overhead of the \EXCESS monitoring framework does not
exceed 3\% assuming that the sampling is not more frequent than one per second.
In turn, the CPU overhead caused by the \ECO Zabbix agents is closer to 10\% if the number of monitoring metrics is
higher than 15. However, a system based on Zabbix also supports virtual layer metrics
in addition to the physical level. 

The monitoring server (MONITOR component) of the \EXCESS monitoring framework is collecting data,
where Elasticsearch database is used in order to store monitoring data.
The \ECO Monitoring component based on a standard Zabbix aggregator relying on
MySQL DB is used in the \ECO monitoring framework, which allows for maintaining more complex
data models with interconnected tables.

Definition of custom metrics is possible within \EXCESS monitoring framework by simply editing a C configuration file.
This is also possible within the \ECO monitoring framework by using the corresponding functionality of Zabbix
and the energy-related values provided by the PDUs.


\begin{figure}[t]
	\centering
	\includegraphics[width=.5\textwidth]{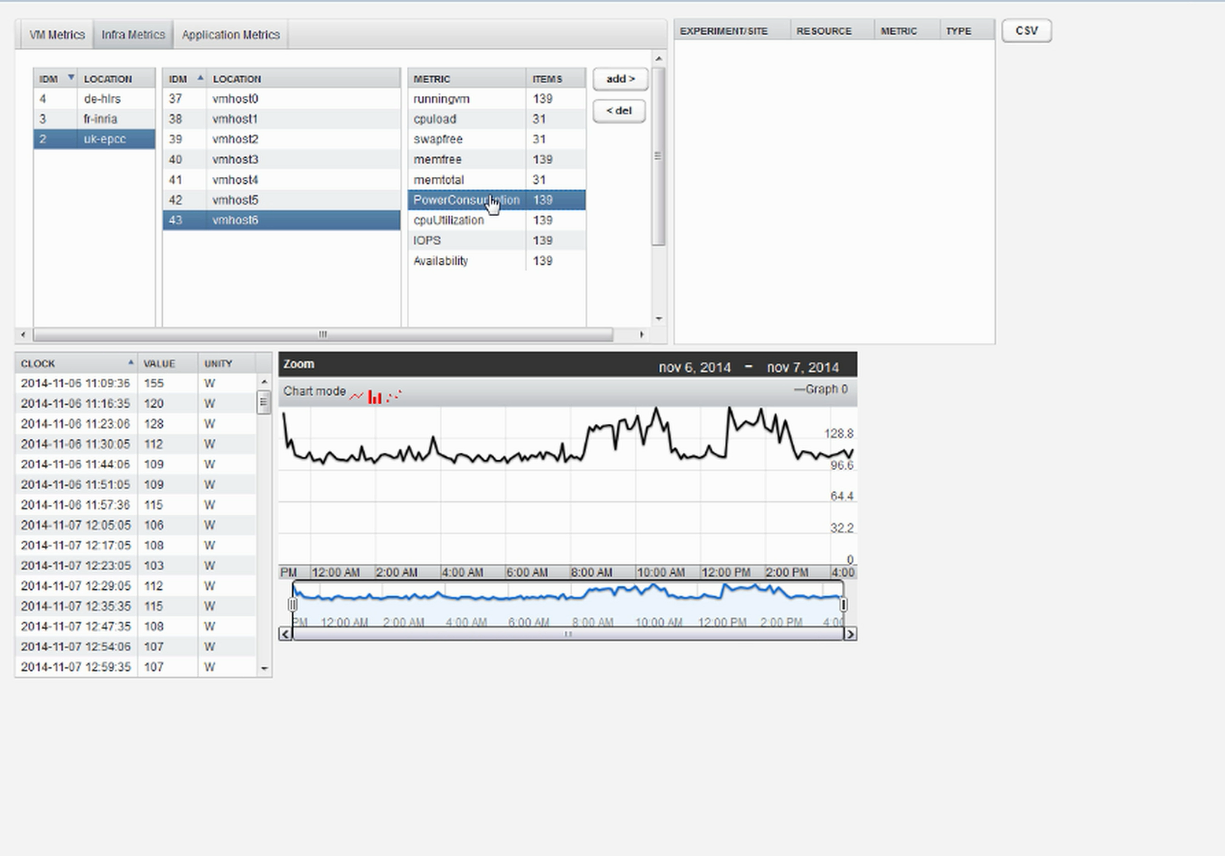}
	\caption{Web-interface of the \ECO portal}
	\label{fig:demo_e2c_monitoring}
\end{figure}

\begin{figure}[t]
	\centering
	\includegraphics[width=.5\textwidth]{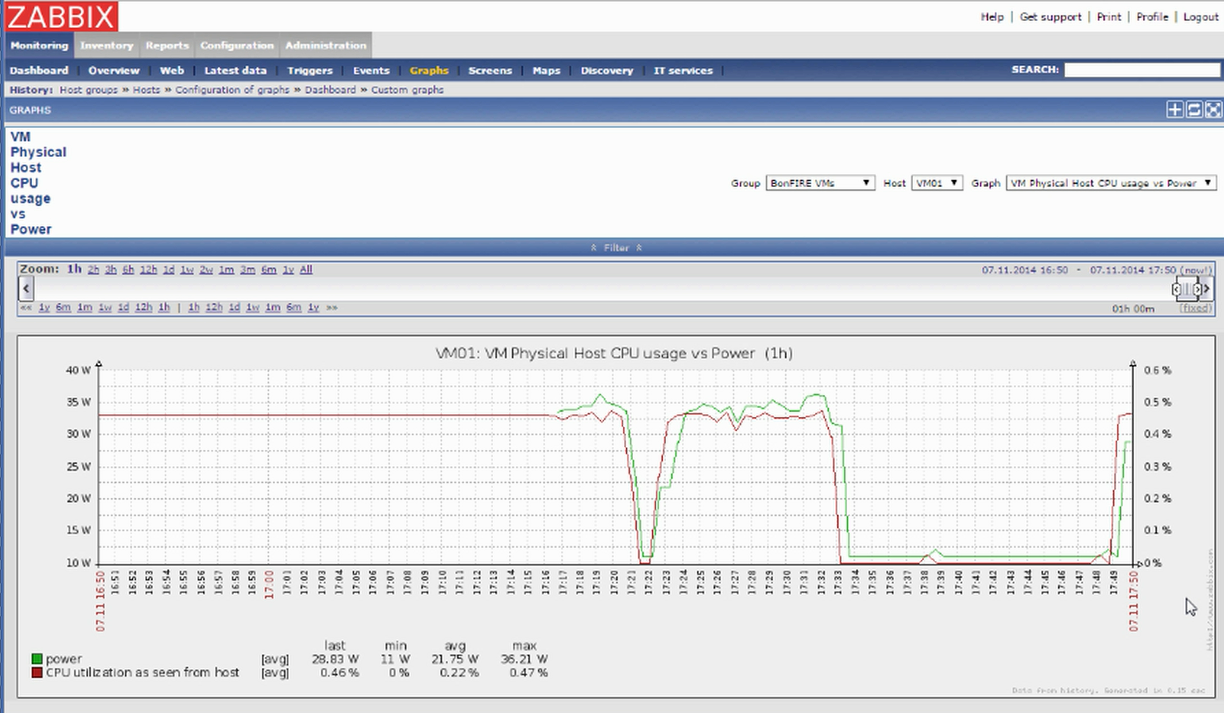}
	\caption{Web-interface of the Zabbix monitoring system used by the \ECO monitoring framework}
	\label{fig:demo_zabbix_monitoring}
\end{figure}

A GUI was developed for the \ECO approach in order to visualize the predicted $\mathrm{CO_2}$
emissions of the application and therefore to help user with choosing the optimal deployment
(cf.~Figure~
\ref{fig:demo_e2c_monitoring}).
In addition, \ECO monitoring framework allows for using the standard Zabbix-based
visualization of metrics (cf.~Figure~\ref{fig:demo_zabbix_monitoring}).

\begin{figure}[t]
	\centering
	\includegraphics[width=.5\textwidth]{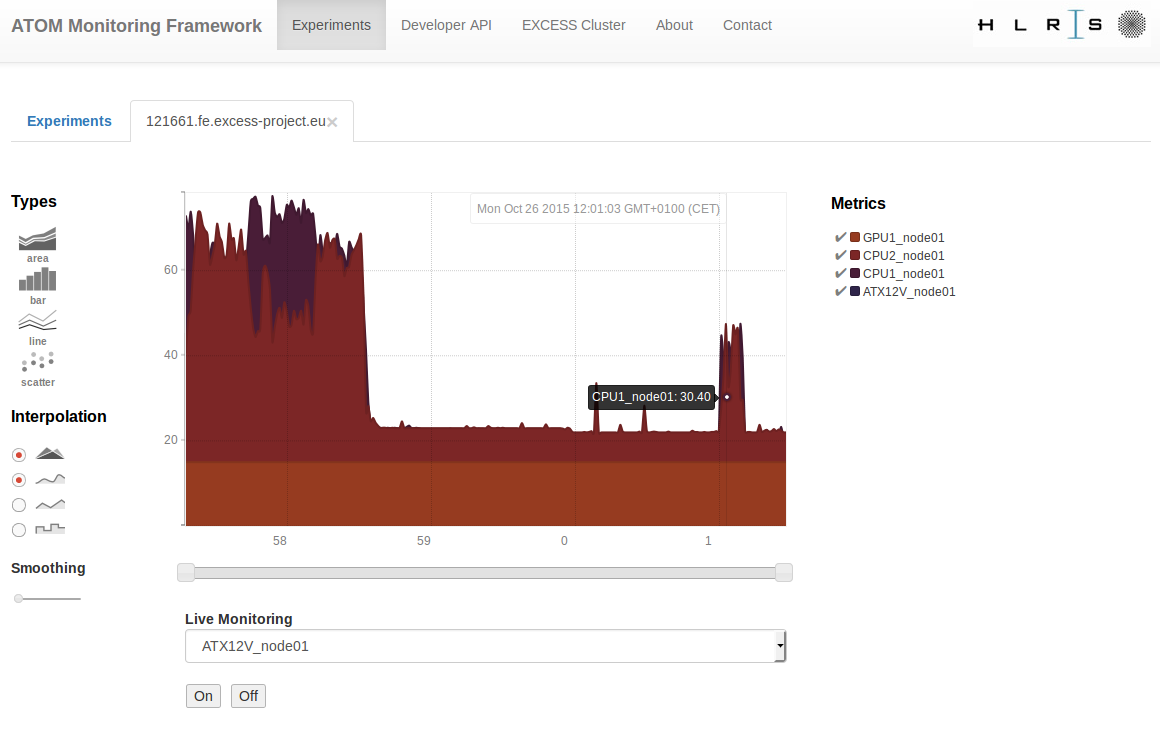}
	\caption{Interactive visualization of running and historic experiments as provided by the web-frontend of the EXCESS monitoring framework}
	\label{fig:excess-diagram}
\end{figure}

A similar GUI was developed on the top of \EXCESS monitoring framework,
which shows a diagram for the specified metric values for the given time frame
(cf.~Figure~
\ref{fig:excess-diagram}).


To summarize our comparison, we consider the key properties which characterize monitoring
frameworks according to the literature \cite{aceto13,katsaros11,telesca14}.
Such properties with respect to the \ECO and \EXCESS monitoring frameworks are listed in Table~\ref{compar}.
We can conclude that the \EXCESS monitoring approach is better suitable for lightweight
monitoring of physical hardware, especially if it is required to be conducted in real time, while
the \ECO monitoring approach is better to be used to collect sophisticated user-defined metrics
on virtual and application level.

\begin{table}
\begin{center}
\caption{Key properties of the \ECO and \EXCESS monitoring frameworks}
\label{compar}
\begin{tabular}{|c|m{2.5cm}|m{2.5cm}|} 
\hline
Key property & \ECO monitoring framework & \EXCESS monitoring framework \\ \hline
Architecture & agent-based & agent-based \\ \hline
Non-intrusiveness & medium & low-intrusive \\ \hline
Scalability & low update frequency & high update frequency \\ \hline
Timeliness & one update per second & near-real time update \\ \hline
Extensibility & extra metrics possible via Zabbix & extra metrics possible via custom plug-ins \\ \hline
Data Storage & MySQL DB & Elasticsearch \\ \hline
Visualization & Zabbix-based & web-based\\ \hline
Adaptability & run-time configuration & run-time configuration \\ \hline
Predictability & no & no \\ \hline
\end{tabular}
\end{center}
\end{table}

%% file: 05-Conclusion.txt
\section{Conclusion}
\label{sec:conclusion}

\noindent
In this paper, we described two alternative solutions for monitoring frameworks originated from the \ECO and
\EXCESS EU projects.

The \EXCESS monitoring framework is well suited for the fine-granular metrics gathering on the physical host level.
Relying on agent-based architecture, it provides low-intrusive real-time monitoring.

In turn, the \ECO monitoring framework provides more advanced energy-aware metrics support on physical,
virtual and application level. It allows for the scheduler to optimize the applications deployment
between cloud provider sites as well as within a single cloud in order to minimize the resulting $\mathrm{CO_2}$ emissions.

\section*{Acknowledgment}
The research leading to these results has received funding from the European Union
Seventh Framework Programme (FP7/2007-2013) under the grant agreements 611183 (EXCESS Project)
and 318048 (\ECO Project).